\documentclass[9pt,shortpaper,twoside]{IEEEtran}
\usepackage{hyperref}
\hypersetup{colorlinks = true,
            linkcolor = red,
            urlcolor  = blue,
            citecolor = blue,
            anchorcolor = blue}
\usepackage{amsmath,amsthm,amssymb,amsfonts}
\usepackage{dsfont}
\usepackage{graphicx}
\usepackage{algorithm,algorithmic}
\usepackage{xcolor}
\usepackage{cite}
\usepackage{subcaption}
\usepackage{epstopdf}
\usepackage{bm}
\usepackage{mleftright}
\usepackage{mathrsfs}
\usepackage{enumerate}
\usepackage{enumitem}

\newtheorem{theorem}{Theorem}
\newtheorem{example}{Example}
\newtheorem{proposition}{Proposition}
\newtheorem{lemma}{Lemma}
\newtheorem{corollary}{Corollary}

\newtheorem{definition}{Definition}

\newtheorem{remark}{Remark}

\newcommand{\field}[1]{\mathbb{#1}}
\newcommand{\R}{\field{R}}
\newcommand{\RP}{\mathcal{R}}
\newcommand{\RH}{\mathcal{RH}_\infty}
\newcommand{\RL}{\mathcal{RL}_\infty}
\newcommand{\C}{\field{C}}

\newcommand{\Z}{\field{Z}}
\newcommand{\esssup}{\mathop{\mathrm{ess\,sup}}}

\newcommand{\bsmtx}{\left[ \begin{smallmatrix}} 
\newcommand{\esmtx}{\end{smallmatrix} \right]}

\newcommand{\TwoOne}[2]
{\begin{bmatrix}
{#1} \\
{#2}
\end{bmatrix}
}

\newcommand{\OneTwo}[2]
{\begin{bmatrix}
{#1} &
{#2}
\end{bmatrix}
}

\newcommand{\mOneTwo}[2]
{\left[ \begin{smallmatrix}
{#1} &
{#2}
\end{smallmatrix} \right]
}
\newcommand{\mTwoOne}[2]
{\left[ \begin{smallmatrix}
{#1} \\
{#2}
\end{smallmatrix} \right]
}

\newcommand{\TwoTwo}[4]
{\begin{bmatrix}
{#1} & {#2} \\
{#3} & {#4}
\end{bmatrix}
}

\newcommand{\mTwoTwo}[4]
{\left[ \begin{smallmatrix}
{#1} & {#2} \\
{#3} & {#4}
\end{smallmatrix}\right]
}

\begin{document}

\title{\LARGE \bf 
On the exponential convergence of input-output signals \\ of nonlinear feedback systems
}

\author{Lanlan Su, Di Zhao, and Sei Zhen Khong % <-this % stops a space
\thanks{*This work was supported in part by the National Key Research and Development Program of China under No. 2022YFA1004700 and by the National Science and Technology Council of Taiwan under No. 113-2222-E-110-002-MY3. Corresponding author: Di Zhao.}% <-this % stops a space
\thanks{Lanlan Su is with the Department of Automatic Control \& Systems Engineering, University of Sheffield, Sheffield, S1 3JD, UK
({\tt\small lanlan.su@sheffield.ac.uk}).}%
\thanks{Di Zhao is with the Department of Control Science and Engineering \&
Shanghai Institute of Intelligent Science and Technology, Tongji University, Shanghai, China ({\tt\small  dzhao925@tongji.edu.cn}).}%
\thanks{Sei Zhen Khong is with the Department of Electrical Engineering, National Sun Yat-sen University, Kaohsiung 80424, Taiwan ({\tt\small szkhong@mail.nsysu.edu.tw}).}
}

\maketitle

\begin{abstract} 
This note studies the exponential convergence of input-output signals of discrete-time nonlinear systems composed of a feedback interconnection of a linear time-invariant system and a nonlinear uncertainty. Both the open-loop subsystems are allowed to be unbounded. 
Integral-quadratic-constraint-based conditions are proposed for these uncertain feedback systems, including the
Lurye type, to exhibit the property that the endogenous input-output signals enjoy an exponential convergence rate for all initial
conditions of the linear time-invariant subsystem. The conditions are established via a combination of tools, including integral quadratic constraints, directed gap, and exponential weightings. 
% We provide conditions under which the feedback system exhibits the property that the endogenous input-output signals enjoy an exponential convergence rate, by using perturbation analysis and a combination of tools including integral quadratic constraints, directed gap, and exponential weightings.  
% As a byproduct, we show that the feedback system with null exogenous inputs  admits such an exponential convergence property for all initial conditions of the linear time-invariant subsystem. }

% We show that the integral-constraint-based robust feedback stability theorem for uncertain nonlinear systems, including the
% Lurye type, exhibits the property that the endogenous input-output signals enjoy an exponential convergence rate for all initial
% conditions of the linear time-invariant subsystem. More generally, we provide conditions under which a feedback interconnection
% of possibly open-loop unbounded subsystems admits such an exponential convergence property, using perturbation analysis
% and a combination of tools including integral quadratic constraints, directed gap, and exponential weightings.
\end{abstract}


\section{Introduction}

\IEEEPARstart{T}{wo} main approaches to nonlinear systems stability analysis are the state-space and input-output methods. The former is dominated by Lyapunov-type theory~\cite{khalil2002nonlinear,van2000l2}. An appealing advantage of the Lyapunov theory is that both asymptotic and exponential stability are readily accommodated. Asymptotic convergence to an equilibrium (or, more generally, a set) often leaves much to be desired, given that the rate of convergence may be arbitrarily slow. On the other hand, exponential convergence is a stronger property than asymptotic convergence if it can be established, since in this case, the user has information on the rate at which the convergence takes place. In the input-state-output  framework~\cite[Chapter 3]{van2000l2}, the notion of exponential dissipativity~\cite{haddad2011nonlinear,khong2022feedback} is valuable when it comes to exploiting Lyapunov theory to establish exponential stability. In numerical analysis~\cite{nocedal1999numerical}, exponential convergence is known as linear convergence, and Lyapunov-type methods have been successfully deployed to show also sub-linear and super-linear convergence.

The input-output approach to nonlinear systems stability analysis, on the contrary, involves showing that the input to the system belonging to some Banach space implies that the output also belongs to a Banach space~\cite{zames1966input,desoer2009feedback,teel1996graphs,teel2010input}, typically taken to be the Lebesgue $L_p$ or $\ell_p$ space. As such, the input-output approach appears lacking in that it can at most guarantee asymptotic convergence in the output. Under some conditions, it is possible to show that global exponential stability implies input-output stability~\cite[Section 6.3]{vidyasagar2002nonlinear}\cite[Section 7.6]{haddad2011nonlinear}, which demonstrates that exponential stability in the state-space framework is a stronger property than input-output stability. On the other hand, the result~\cite[Proposition 1]{megretski1997system} establishes exponential convergence to the origin from the input-output stability of a Lurye system, under an additional boundedness condition on the static nonlinearity and the requirement that the linear time-invariant (LTI) component is exponentially stable. This renders it inapplicable to certain static monotone nonlinear functions or convex optimisation algorithms that are marginally/critically stable. Interestingly, \cite[Section 5]{rantzer1997system} shows that when both the open-loop systems have fading memory, the closed-loop system's input-output stability implies exponential stability. Nonetheless, in the context of robust exponential stability, i.e. exponential convergence in feedback systems in the presence of uncertainty, it supplies no a priori bound on the exponential convergence rate over the set of uncertainty, and the convergence rate may be arbitrarily slow if the set is open.

{The main objective} of this paper is to provide conditions that guarantee exponential convergence of the endogenous $\ell_2$ input-output signals in feedback interconnections of possibly open-loop unbounded {discrete-time} subsystems, within the powerful and unifying systems analysis framework based on integral quadratic constraints (IQCs)~\cite{megretski1997system,rantzer1997system,khong2021integral}. {The proofs of the main results make} use of perturbation analysis with respect to a directed gap introduced in~\cite{georgiou1997robustness} as well as the exponential weighting methods in~\cite{lessard2016analysis,hu2016exponential}. { IQC-based perturbation analysis with respect to a directed gap was first introduced in \cite{rantzer1997system} for nonlinear feedback stability analysis, and related results based on the $\nu$-gap metric have been developed  in~\cite{cantoni2011robustness} for the LTI setting and~\cite{cantoni2013robust} for the linear time-varying setting; all of these works were concerned with feedback stability as opposed to exponential convergence in the present paper.}  The main results in this paper are subsequently specialised to Lurye systems involving static and time-invariant nonlinearity. In this case, a uniform bound on the exponential convergence rate over the uncertainty set is shown to be obtainable from our result in the form of a linear matrix inequality (LMI) condition. 

Among the relevant works, it is noteworthy that~\cite{lessard2016analysis,hu2016exponential} employ IQCs in the hard (a.k.a. unconditional) form~\cite{megretski2010integral}, wherein the integrals are taken over all finite times for signals in extended spaces, much in the spirit of~\cite{seiler2014stability} and the classical small-gain and passivity theorems~\cite{zames1966input,desoer2009feedback,van2000l2}. One major shortcoming in doing so is that it excludes the utility of several highly useful soft (a.k.a. conditional) IQCs, where the integrals are taken from $0$ to $\infty$ for square-integrable signals, such as those involving general forms of Zames--Falb~\cite{Zames68,carrasco2013equivalence,carrasco2016zames,khong2021necessity} and Willems--Brockett multipliers~\cite{Willems68,su2023,zhang2022zames}, as well as the Popov multipliers~\cite{megretski1997system,jonsson1997stability}. On the contrary, the use of hard IQCs provides the advantage of bounding the transient behaviour of the signals, in addition to establishing the exponential convergence, which concerns the tail behaviour. This sets~\cite{lessard2016analysis,hu2016exponential} apart from the results in this paper, and the related work~\cite{michalowsky2021robust} that employs soft IQCs in conjunction with loop transformations. Nevertheless, as far as analysing robust closed-loop stability is concerned, the fusion of soft IQCs and gap metric has been shown to be more powerful than using hard IQCs alone~\cite{khong2021integral}. 

 A directed gap is used as a mathematical tool throughout this technical note to accommodate open-loop unbounded systems in the spirit of~\cite{rantzer1997system,cantoni2013robust,khong2021integral,dizhao2020tac,dizhao2021auto}. An alternative would involve employing loop transformations to encapsulate the unbounded component in an artificial feedback loop that defines a bounded operator and deriving novel IQCs for the latter, as is done in~\cite{michalowsky2021robust,jonsson2000zames,megretski2001new}. It is worth noting that the approach taken in this work is more direct and general, since the aforementioned loop transformation method { is established for specific instances (e.g. for monotone or quasi-concave static nonlinearities in~\cite{michalowsky2021robust,jonsson2000zames,megretski2001new}). Besides, loop transformations have been known to be subsumable by the theory of IQCs; see~\cite[Proposition 5]{jonsson2000lecture}.}
 % and for different types of systems, the loop transformations need to be distinctly designed. Moreover, under some circumstances, the loop transformation may potentially destroy the inherent structure --- such as sparsity --- in the open-loop systems.
{The main contributions of this work are summarised as follows: 
\begin{itemize}
    \item  We propose conditions to establish robust exponential stability for a general class of nonlinear feedback systems, consisting of a possibly unbounded LTI system and a possibly unbounded nonlinear system belonging to some uncertainty set.
\item We provide an LMI characterisation of the problem of optimising the uniform bound on the exponential convergence rate across the uncertainty set.
\item We tailor our results to Lurye systems by simplifying the verification of robust exponential stability conditions when the uncertain nonlinear open-loop system is static and bounded.
\end{itemize}}

The remainder of the paper is organised as follows. The next section introduces the notation and certain definitions used throughout the paper. 
The main results are presented in Section \ref{sec: main}. {Specifically, Section  \ref{subsec:decay rate} provides conditions for establishing the robust exponential convergence of input-output signals of the nonlinear feedback systems, Section~\ref{sec: LMI} an LMI characterisation of the condition on the LTI part of the feedback system, and Section~\ref{sec: ex} numerical examples illustrating the utility of the main results. The exponential stability results are subsequently specialised to Lurye systems in Section \ref{subsec: Lurye}, where the uncertain nonlinearity is restricted to being time-invariant, static (a.k.a. memoryless), and bounded.} Finally, some concluding remarks are provided in Section~\ref{sec: con}.

\section{Notation and preliminaries}
Let $\R$, $\C$, $\Z$, $\Z_{0}^{+}$, $\Z^{-}$ denote the sets of real numbers, complex numbers, integers, non-negative integers, and negative integers respectively. Denote by $\|v\|$ the Euclidean norm of a vector $v\in\R^{n}$. {Denote by $A^\top$ and $A^*$ the transpose and complex conjugate transpose of a matrix $A\in\C^{n\times n}$, respectively.}  
Let $\lambda_1,\ldots,\lambda_n$ be the eigenvalues of a matrix $A\in\C^{n\times n}$, and the spectral radius of $A$ is denoted by $r(A)=\max\{|\lambda_1|,\ldots,|\lambda_n|\}$. If $r(A) < 1$, $A$ is said to be Schur stable. Denote by $\bar{\sigma}(B)=\sigma_1(B)\geq\cdots\geq\sigma_n(B)=\underline{\sigma}(B)$ the singular values of $B\in\mathbb{C}^{p\times q}$ with $n=\min\{p,q\}$. {Denote by $I$ the identity matrix.}

Let $\ell_2$ be the set of (two-sided) discrete-time signals $u:\Z \to \R^n$ satisfying
$\sum_{k\in\Z} u_k^{\top} u_k <\infty$.
This forms an Hilbert space with the inner product $\langle u,w\rangle := \sum_{k\in\mathbb{Z}}u_k^{\top} w_k$ and the
corresponding norm $\|u\|:=\langle u,u \rangle^{1/2}$.
Also define the one-sided
signal space $\ell_2^{0+} := \left\{ f \in \ell_2 :  f_i=0, \forall i<0 \right\}$. For any discrete-time signal $u$ and  $\tau\in \Z_0^+$, define
the truncation operator $P_\tau$
by
$(P_\tau u)_k:=u_k$ for $ k \le \tau$ and $(P_\tau u)_k:=0$ for $k> \tau$.  The one-sided extended space
is defined as $\ell_{2e}^{0+}  := \left\lbrace f: \mathbb{Z} \rightarrow \mathbb{R}^n : P_\tau f\in \ell_2^{0+} ,\forall \tau\in \mathbb{Z}_0^+ \right\rbrace$. 

% The shift operation is defined by $(S_\tau u)_k:=u_{k-\tau}$. 

\subsection{Open-loop systems}
A system $\Gamma:\ell_{2e}^{0+}\rightarrow\ell_{2e}^{0+}$ is said to be \emph{causal} if $P_\tau \Gamma P_\tau =P_\tau \Gamma$ for all $\tau\in\Z_0^{+}$.  A causal system $\Gamma:\ell_{2e}^{0+}\rightarrow\ell_{2e}^{0+}$ is said to be \emph{bounded} (or \emph{stable}) if 
\begin{align*}
    \|\Gamma\|:= \sup_{ \tau \in\Z^+_0;0 \neq P_\tau u \in \ell_{2e}^{0+}}
    \frac{\|P_\tau \Gamma u\|}{\|P_\tau u\|}=\sup_{0\neq u \in \ell_2^{0+}}
    \frac{\|\Gamma u\|}{\|u\|} <\infty,
\end{align*}
where the second equality follows from~\cite[Theorem 2.1]{Willems1971nonlinear}. 
Given a causal system $\Gamma:\ell_{2e}^{0+}\rightarrow \ell_{2e}^{0+}$, define the graph of $\Gamma$ as $\mathscr{G}(\Gamma):=\left\{\mTwoOne{x}{y}\in\ell_2^{0+}\;:\;y=\Gamma x\right\}$, 
and the inverse graph as $\mathscr{G}'(\Gamma):=\left\{\mTwoOne{x}{y}\in\ell_2^{0+}\;:\;x=\Gamma y\right\}$. 
Similarly, define the extended graph of $\Gamma$ as $\mathscr{G}_e(\Gamma):=\left\{\mTwoOne{x}{y}\in\ell_{2e}^{0+}\;:\;y=\Gamma x\right\}$, and the inverse extended graph of $\Gamma$ as $\mathscr{G}_e'(\Gamma):=\left\{\mTwoOne{x}{y}\in\ell_{2e}^{0+}\;:\;x=\Gamma y\right\}$.
Let $\vec{\delta}(\cdot,\cdot)$ denote the \emph{directed gap} \cite{georgiou1997robustness} between two (nonlinear) causal systems $\Gamma_1$ and $\Gamma_2$, which is defined as
\begin{align}\label{eq:direct_gap}
\vec{\delta}(\Gamma_1,\Gamma_2):=\limsup_{\tau\to\infty}\hspace{-2pt}\sup_{v\in\mathscr{G}_e (\Gamma_2)}\inf_{\substack{u\in\mathscr{G}_e (\Gamma_1),\\ \|P_\tau u\|\neq 0}}\frac{\|P_\tau(u-v)\|}{\|P_\tau u\|}.
\end{align}
Given a bounded interval $\Lambda\subset\mathbb{\R}$, a mapping $\lambda\in \Lambda\mapsto \Gamma_\lambda$ is said to be continuous in  {(the directed)} gap if for any $\epsilon>0$, there exists $\delta>0$ such that for all $\nu,\mu\in\Lambda$ with $|\nu-\mu|<\delta$ it holds
$\vec{\delta}(\Gamma_{\nu},\Gamma_{\mu})<\epsilon.$
We say that a set of nonlinear systems $\mathbf{\Gamma}$ is path-connected in gap from $\Delta$ if $\Delta\in\mathbf{\Gamma}$  and for any $\bar{\Gamma}\in\mathbf{\Gamma}$ there exists a mapping $\lambda\in[a,b]\mapsto\Gamma_\lambda\in\mathbf{\Gamma}$, $a<b$, that is continuous in gap with $(\Gamma_a,\Gamma_b)=(\Delta,\bar{\Gamma})$ or $(\Gamma_b,\Gamma_a)=(\Delta,\bar{\Gamma})$. 

%In the above definitions, it is noteworthy that the order of arguments in %$\vec{\delta}(\Gamma_{\nu},\Gamma_{\mu})$ can be switched as $\nu$ and $\mu$ %are not ordered, in which case $(\Gamma_a,\Gamma_b)=(\Delta,\bar{\Gamma})$ %can also be equivalently set as $(\Gamma_b,\Gamma_a)=(\Delta,\bar{\Gamma})$. 

A causal system $\Gamma:\ell_{2e}^{0+}\rightarrow \ell_{2e}^{0+}$ is said to be locally Lipschitz continuous if 
$$\sup_{\substack{x,y\in\ell_{2e}^{0+},\\ P_\tau(x-y)\neq 0}}\frac{\|P_\tau(\Gamma x-\Gamma y)\|}{\|P_\tau(x-y)\|}<\infty,~\forall\tau\in\mathbb{Z}_0^+.$$
When this is the case, its uniform instantaneous gain \cite[Section~4.3.3]{Willems1971nonlinear} is defined as
$$\gamma(\Gamma):=\sup_{\substack{x,y\in\ell_{2e}^{0+},\tau\in\mathbb{Z}_0^+,\\ P_\tau(x-y)= 0,P_{\tau+1}(x-y)\neq 0}}\frac{\|P_{\tau+1}(\Gamma x-\Gamma y)\|}{\|P_{\tau+1}(x-y)\|}.$$
 {Let $\mathcal{L}_\infty$ denote the set of complex functions $\hat{M}:\C\to\C^{n\times n}$ (a.e.) that are  even symmetric and essentially bounded on the unit circle, i.e., $\hat{M}(e^{-j\omega}) = \overline{\hat{M}(e^{j\omega})}$, $\omega\in[0,2\pi)$ and
\[
 \|\hat{M}\|_\infty:=\esssup_{\omega\in[0,2\pi)}\bar{\sigma}(\hat{M}(e^{j\omega}))<\infty.
\]}%
 {Every bounded linear time-invariant (LTI) system mapping $\ell_2$ to $\ell_2$ is associated with an element in $\mathcal{L}_\infty$.} Specifically, $M:\ell_2\rightarrow\ell_2$ 
is bounded and LTI if and only if it admits a transfer
function $\hat{M} \in \mathcal{L}_\infty$
such that $y=Mu$ is equivalent via the discrete-time Fourier transform to
multiplication by $\hat{M}$ in the frequency domain:
$\hat{y}(e^{j\omega}) =\hat{M}(e^{j\omega}) \hat{u}(e^{j\omega})$, where $\hat{u}(e^{j\omega}) := \sum_{n\in \Z} u_n e^{-j \omega n}$ and $\hat{y}(e^{j\omega}) := \sum_{n\in \Z} y_n e^{-j \omega n}$.  
Let $\RP$ denote the set of proper real-rational transfer functions, $\mathcal{RL}_\infty:= \RP\cap \mathcal{L}_{\infty}$ denote the space of proper real-rational transfer functions with no poles on the unit circle, and
$\mathcal{RH}_\infty$ denote the space of proper real-rational transfer functions with all poles in the open unit disk. Every $G\in \mathcal{RH}_\infty$ is associated with a causal bounded LTI system $G:\ell_{2e}^{0+}\rightarrow\ell_{2e}^{0+}$ satisfying $\|G\| = \|\hat{G}\|_\infty$. We do not differentiate between the abovementioned frequency-domain and time-domain representations for notational convenience. Given $\Pi\in \mathcal{L}_{\infty}$, the adjoint of $\Pi$ is defined as the operator $\Pi^*\in\mathcal{L}_{\infty}$ such that $\langle u, \Pi^* y\rangle=\langle \Pi u,y\rangle$ for all $u,y\in\ell_2$. An operator in $\mathcal{L}_\infty$ is said to be self-adjoint if $\Pi=\Pi^*$. For any transfer function $G\in\RP$, we use $G\sim(A,B,C,D)$ to denote that $G$ admits the realisation $(A,B,C,D)$, i.e. $G(z) = C(zI - A)^{-1}B + D$. A pair $(N,M)$ is said to be a right coprime factorisation (RCF) of $G \in \RP$ if $N,M\in\RH$, $G=NM^{-1}$ and $(N, M)$ are coprime over $\RH$, i.e. there exist $X, Y \in \RH$ such that $XN + YM = I$; see, for instance, \cite[Section 1.2.1]{vinnicombe2001}. Note that an RCF always exists for any $G \in \RP$.
\subsection{Closed-loop systems}
Consider $G,\Delta:\ell_{2e}^{0+}\rightarrow\ell_{2e}^{0+}$ and throughout this work, we assume both $G$ and $\Delta$ are causal, locally Lipschitz continuous, and map $0$ to $0$.  

\begin{figure}
\centering
 \includegraphics[scale=0.3]{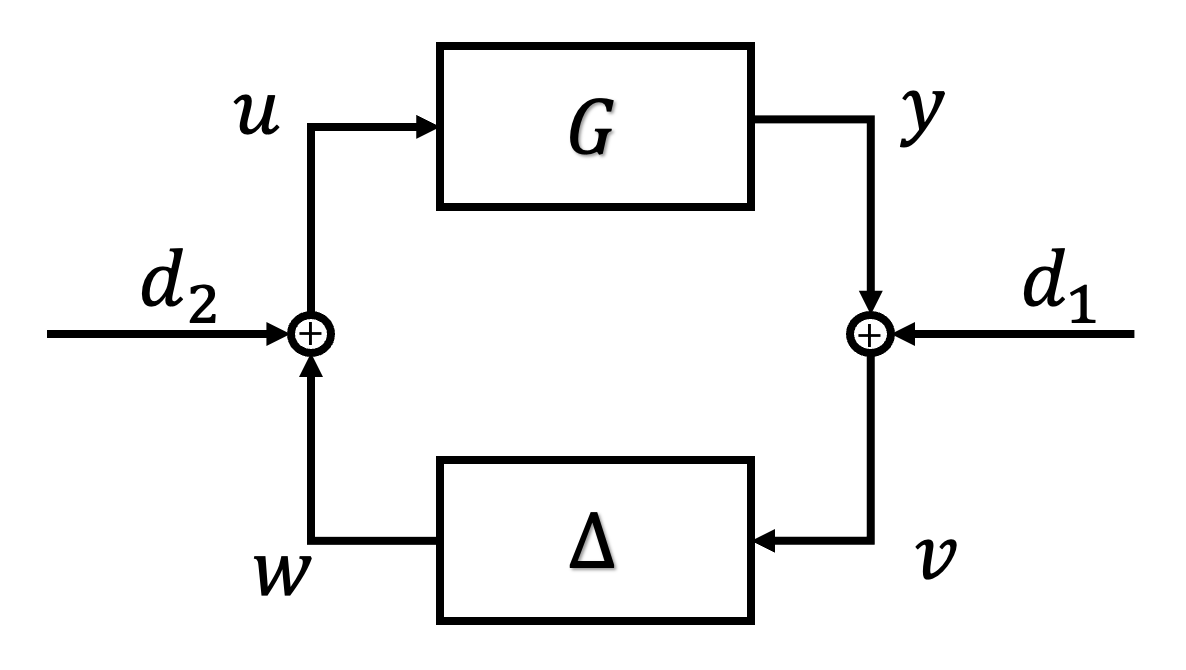}  
 \caption{Feedback system}
 \label{fig:lure}
\end{figure}

\begin{definition}\label{def:wp}{\rm 
 (\cite{rantzer1997system})}
 The feedback system $[G,\Delta]$ illustrated in Fig.~\ref{fig:lure} is said to be \underline{well-posed} if  for any (exogenous) signals $d_1,d_2\in \ell_{2e}^{0+}$ there exist endogenous signals $u,y,v,w\in \ell_{2e}^{0+}$ that satisfy 
\begin{align}\label{eq:syst_eq}
    y  =Gu;\quad  
    w  =\Delta v;\quad
    v  =y+d_1;\quad
    u  =w +d_2,
\end{align}
and depend causally on $d_1,d_2$.
\end{definition}

% {A  stricter version of well-posedness has been defined in \cite[Chpater 4] {Willems1971nonlinear} wherein the insensitivity to modeling errors is required.
% The following lemma tailored from \cite[Theorem~4.1]{Willems1971nonlinear} gives a sufficient condition for $[G,\Delta]$ to be well-posed as per the definition therein, and thus as per Definition \ref{def:wp}.} 
\begin{lemma}[{\cite[Chap.~4]{Willems1971nonlinear}}]\label{lem:wellposed_inst}
For locally Lipschitz continuous $G,\Delta:\ell_{2e}^{0+}\rightarrow\ell_{2e}^{0+}$, $[G,\Delta]$ is well-posed if
$\gamma(G)\gamma(\Delta)<1$.
\end{lemma}
Let $\mathbf{\Delta}$ be a class of causal nonlinearities. 
\begin{definition}\label{def:robust stability}(Adapted from \cite{rantzer1997system})
 $[G,\Delta]$ is said to be 
  \underline{robustly stable} over $\mathbf{\Delta}$
  if $[G,\Delta]$ is well-posed for all
  $\Delta \in \mathbf{\Delta}$ and for every $\Delta \in \mathbf{\Delta}$ there exists
  $\kappa>0$ such that 
\begin{multline*}
\sum_{k=0}^{\tau} \|u_k\|^2+\|y_k\|^2+\|v_k\|^2+\|w_k\|^2\le \\
\kappa \sum_{k=0}^{\tau} \left(\|d_{1k}\|^2+ \|d_{2k}\|^2\right)
\end{multline*}
for all $\tau\in\Z_0^{+}$ and $d_1,d_2\in\ell_{2e}^{0+}$.
\end{definition}

\section{Main results} \label{sec: main}
% In this work, we consider the robust stability of the Lurye system $[G, \Delta]$ depicted in Fig.~\ref{fig:lure} where   $G: \ell_{2e}^{0+}\rightarrow\ell_{2e}^{0+}$  is a causal LTI system, and   $\Delta:\ell_{2e}^{0+}\rightarrow\ell_{2e}^{0+}$ is a causal nonlinearity. 

We formulate in this section the problem addressed in this work. We begin by introducing some definitions and then present a key preliminary result.

Define the following exponentially scaling operators 
$\rho_{+},\rho_{-} : \ell_{2e}^{0+}\rightarrow\ell_{2e}^{0+}$, for $\rho\in(0,1]$, via $(\rho_{+}(y))_k=\rho^{k}y_k,k\in\Z_0^+$ and   $(\rho_{-}(y))_k=\rho^{-k}y_k, k\in\Z_0^+$, respectively. Evidently, $\rho_{+},\rho_{-}$ are the inverse of  each other. Note that if $\rho_-y\in\ell_2^{0+}$, then $\lim_{k\rightarrow \infty}\rho^{-k}y_k=0$, i.e. $y\in\ell_{2}^{0+}$ decays exponentially at a rate at least as fast as $\rho$. That is, for every $\rho_- y\in\ell_{2}^{0+}$, there exists $\eta > 0$ such that
$
\|y_k\|\le \eta \hat{\rho}^{k}
$
for all $\hat{\rho} \in [\rho, 1)$ and $k \in \mathbb{Z}_0^+$.
With the above understanding, we define the notion of robust exponential stability in what follows. Let $\mathbf{\Delta}$ be a class of causal nonlinearities.

\begin{definition}\label{def:robust exponential stability}
 $[G,\Delta]$ is said to be \underline{robustly exponentially stable} over $\mathbf{\Delta}$
  if $[G,\Delta]$ is robustly stable over $\mathbf{\Delta}$, and for some $\rho\in(0,1)$ and every $\Delta\in\mathbf{\Delta}$, $d_1,d_2\in\rho_+\ell_2^{0+}$, there exists
  $\eta>0$ such that 
\begin{align*}
\|y_k\|\le \eta \rho^{k} \text{  and  } \|w_k\|\le \eta\rho^{k}, \;\forall k\in\Z_0^{+}.
\end{align*}
In the case where such $\rho$ is known, $[G,\Delta]$ is robustly exponentially stable over $\mathbf{\Delta}$ with rate at most $\rho$, and $\rho$ is called an \emph{exponential decay rate}. The smaller $\rho$  is, the faster the endogenous signals decay. 
\end{definition}

\begin{remark}
The $\eta$ in Definition~\ref{def:robust exponential stability} may vary with $\Delta$, $d_1$, and $d_2$. As a consequence, it may vary with different $y$ and $w$. This contrasts with~\cite{lessard2016analysis,hu2016exponential}, where hard IQCs are used and the `$\eta$' therein is uniform in all uncertainties and initial conditions. The difference arises due to the fact that hard IQCs are stronger conditions { (involving summations taken over all finite time intervals starting from 0 for $\ell_{2e}^{0+}$ signals)} than their soft counterparts (where the summations are taken from $0$ to $\infty$ for $\ell_2^{0+}$ signals), and hence give rise to a stronger convergence with uniformity of $\eta$. The downside, nevertheless, is that there exist powerful soft IQCs that do not hold in the hard form such as the Popov IQC ~\cite[Section IV]{megretski1997system}~ and Zames--Falb IQC~\cite[Lemma 8]{Zames68}. A condition that allows transitioning from a soft IQC to a hard one is the existence of a canonical factorisation for the multiplier~\cite{seiler2014stability,carrasco2019conditions}.
\end{remark}

Soft IQCs are employed in this work, and the following soft IQC-based result, adopted from Theorem~2 in \cite{rantzer1997system}, plays a crucial role in this work; see also Theorem IV.2 in \cite{khong2021integral} and the references therein for variants of this result.

\begin{proposition}\label{prop:SZ_IQC}
Suppose that
\begin{enumerate}\renewcommand{\theenumi}{\textup{(\roman{enumi})}}\renewcommand{\labelenumi}{\theenumi}
    \item the mappings $\lambda\in[\underline{\lambda},\bar{\lambda}]\mapsto G_\lambda$ and $\lambda\in[\underline{\lambda},\bar{\lambda}]\mapsto \Delta_\lambda $ are continuous in $\vec{\delta}(\cdot,\cdot)$;
    \item $[G_{\underline{\lambda}},\Delta_{\underline{\lambda}}]$ or $[G_{\bar{\lambda}},\Delta_{\bar{\lambda}}]$ is stable;
    \item $[G_\lambda,\Delta_\lambda]$ is well-posed for all $\lambda\in[\underline{\lambda},\bar{\lambda}]$;
    \item there exists a multiplier $\Pi=\Pi^*\in\mathcal{L}_\infty$ and $\epsilon>0$ such that  for every $\lambda\in[\underline{\lambda},\bar{\lambda}]$, 
    \begin{align} \label{eq:IQC_Delta}
        \Bigg\langle \TwoOne{v}{w}, \Pi \TwoOne{v}{w}\Bigg\rangle\ge 0, \forall \TwoOne{v}{w}\in\mathscr{G}(\Delta_\lambda)
      \end{align}
      and
     \begin{align*}%\label{eq:IQC_G}
        \Bigg\langle \TwoOne{y}{u}, \Pi \TwoOne{y}{u}\Bigg\rangle \le -\epsilon \left\| \TwoOne{y}{u}\right\|^2, \forall \TwoOne{y}{u}\in\mathscr{G}'(G_\lambda).
      \end{align*}
\end{enumerate}
Then, $[G_\lambda,\Delta_\lambda]$ is stable for all $\lambda\in[\underline{\lambda},\bar{\lambda}]$.
\end{proposition}

\subsection{Robust exponential stability}\label{subsec:decay rate}

In this subsection, the main theorem providing robust exponential stability conditions
% that guarantee a specific given exponential decay rate, 
for feedback interconnections of possibly open-loop unbounded systems is presented. Subsequently, the search for an upper bound on the smallest exponential decay rate based on the proposed condition on the LTI subsystem is shown to be transformable into an equivalent LMI.

For a possibly nonlinear operator $\Delta:\ell_{2e}^{0+}\rightarrow \ell_{2e}^{0+}$, define its exponentially weighted operator by $\Delta_\rho = \rho_- \Delta \rho_+$. Given a set of operators $\mathbf{\Delta}$ and a $\rho\in(0,1]$, let $\mathbf{\Delta}_\rho = \{\Delta_\rho:~\Delta\in\mathbf{\Delta}\}$. {Similarly to $\Delta_\rho$, define $G_\rho:=\rho_-G\rho_+$ for $G \in \RP$.}

%define $\lambda_+, \lambda_-: \ell_{2e}^{0+}\rightarrow\ell_{2e}^{0+}$ as $(\lambda_{+}(y))_k=\lambda^{k}y_k,k\in\Z_0^+$ and   $(\lambda_{-}(y))_k=\lambda^{-k}y_k, k\in\Z_0^+$, respectively. For $G \in \RP$ and $\lambda >0$, define $G_\lambda:=\lambda_-G\lambda_+$.}\textcolor{red}{Do we still need this?}

{The following theorem is our first main result in the paper. It provides conditions that guarantee a given exponential decay rate $\rho$ for the endogenous signals in a feedback interconnection of possibly open-loop unbounded systems in Fig.~\ref{fig:lure}.}

\begin{theorem}\label{th:fix rate}
Consider the feedback system in Fig.~\ref{fig:lure} with $G,\Delta:\ell_{2e}^{0+}\rightarrow \ell_{2e}^{0+}$, $G \in \RP$, and $\Delta\in\mathbf{\Delta}$. 
Given  $\rho\in(0,1)$, the feedback system $[G,\Delta]$ is  robustly exponentially stable over $\Delta\in\mathbf{\Delta}$ with decay rate $\rho$ if the following conditions hold:
\begin{enumerate} \renewcommand{\theenumi}{\textup{(\roman{enumi})}}\renewcommand{\labelenumi}{\theenumi}
        \item \label{con:rho_Hstable} $[G_{\rho},H]$ is stable for some $H:\ell_{2e}^{0+}\rightarrow \ell_{2e}^{0+}$ that belongs to $\mathbf{\Delta}_\rho$;
        \item \label{con:rho_homo_Delta}  $\mathbf{\Delta}_\rho$ is path-connected in the directed gap from $H$;
        \item \label{con:rho_wellpose}  $[G,\Delta]$ is well-posed for all $\Delta\in\mathbf{\Delta}$;
        \item \label{con:rho_IQC couple} there exists $\Pi=\Pi^*\in\mathcal{L}_\infty$ such that  
    {\begin{align}\label{eq:rho_IQC on Delta}
        \Bigg\langle \TwoOne{\tilde{v}}{\tilde{w}}, \Pi \TwoOne{\tilde{v}}{\tilde{w}}\Bigg\rangle\ge 0, \forall \TwoOne{\tilde{v}}{\tilde{w}}\in\mathscr{G}(\Delta_\rho), \forall \Delta\in\mathbf{\Delta}
      \end{align}}
and 
\begin{align}\label{eq:FDI_Grho}   \TwoOne{N_{{\rho}}(e^{j\omega})}{M_{{\rho}}(e^{j\omega})}^{*}\Pi(e^{j\omega})\TwoOne{N_{{\rho}}(e^{j\omega})}{M_{{\rho}}(e^{j\omega})}<0, \forall \omega\in[0,2\pi]
\end{align}
\end{enumerate}
where $(N_{\rho},M_{\rho})$ is an RCF of $G_{\rho}$.  
\end{theorem}
Before proving Theorem~\ref{th:fix rate}, we first present in what follows two supporting lemmas. 
The first lemma establishes the equivalence between an IQC and a frequency-domain inequality (FDI). 
\begin{lemma}\label{lem:graph_cf}
Let $\Pi=\Pi^{*}\in\mathcal{L}_{\infty}$, $G \in \RP$, and $(N,M)$ be an RCF of $G$, then the following are equivalent:
\begin{enumerate}\renewcommand{\theenumi}{\textup{(\roman{enumi})}}\renewcommand{\labelenumi}{\theenumi}
    \item \label{con: IQC_l2} there exists   $\epsilon>0$ such that \begin{align*}
       \langle x,\Pi x\rangle\le -\epsilon \left\|x\right\|^2, \forall x \in\mathscr{G}' (G);
    \end{align*}
    \item  \label{con: FDI} it holds that \begin{align*}
        \TwoOne{N(e^{j\omega})}{M(e^{j\omega})}^{*}\Pi(e^{j\omega})\TwoOne{N(e^{j\omega})}{M(e^{j\omega})}<0, \forall \omega\in[0,2\pi].
    \end{align*}
\end{enumerate}
\end{lemma}

\begin{proof}
Firstly, note from \cite[Proposition~1.33]{vinnicombe2001} that 
% the set $\mathscr{G}(G)$ is equivalent to 
% \begin{align*}
%     \left\{\TwoOne{M}{N} q \in \ell_2^{0+}\;:\;q\in \ell_2^{0+} \right\}. 
% \end{align*}
% This means that 
the set $\mathscr{G}'(G)$ is $\left\{\mTwoOne{N}{M} q \in \ell_2^{0+}\;:\;q\in \ell_2^{0+} \right\}$.
Hence, the condition in \ref{con: IQC_l2} can be rewritten into there exists  $\epsilon>0$ such that
\begin{align*}
    \Bigg\langle \TwoOne{N}{M} q,(\Pi+\epsilon I) \TwoOne{N}{M} q\Bigg\rangle\le 0, \forall q\in\ell_2^{0+}.
\end{align*}
This is then equivalent, via the argument in  \cite[Theorem~3.1]{MegTre93}, to 
\begin{align*}
    \Bigg\langle \TwoOne{N}{M} q,(\Pi+\epsilon I) \TwoOne{N}{M} q\Bigg\rangle\le 0, \forall q\in\ell_2.
\end{align*}
Since $M,N,\Pi\in \mathcal{L}_{\infty}$, the above condition is equivalent to the FDI
\begin{align*}
\TwoOne{N(e^{i\omega})}{M(e^{j\omega})}^{*}(\Pi(e^{j\omega})+\epsilon I) \TwoOne{N(e^{j\omega})}{M(e^{j\omega})} \leq 0, \forall \omega\in[0,2\pi].
\end{align*}
This is equivalent to \ref{con: FDI} since $N, M \in \RH$ are coprime over $\RH$.\qedhere
%  {By the fact that $N(e^{i\omega})$ and $M(e^{i\omega})$ are coprime whereby $N(e^{i\omega})^*N(e^{i\omega})+M(e^{i\omega})^*M(e^{i\omega})$ is nonsingular,} we obtain that  condition \ref{con: FDI}  is implied by the above condition. The converse can be proved by observing  that the inequality over the compact set $[0,2\pi]$ in \ref{con: FDI} is strict and the fact that $N, M$ are in $\RH$. 
\end{proof}

\begin{figure}
    \centering
    \includegraphics[scale=0.2]{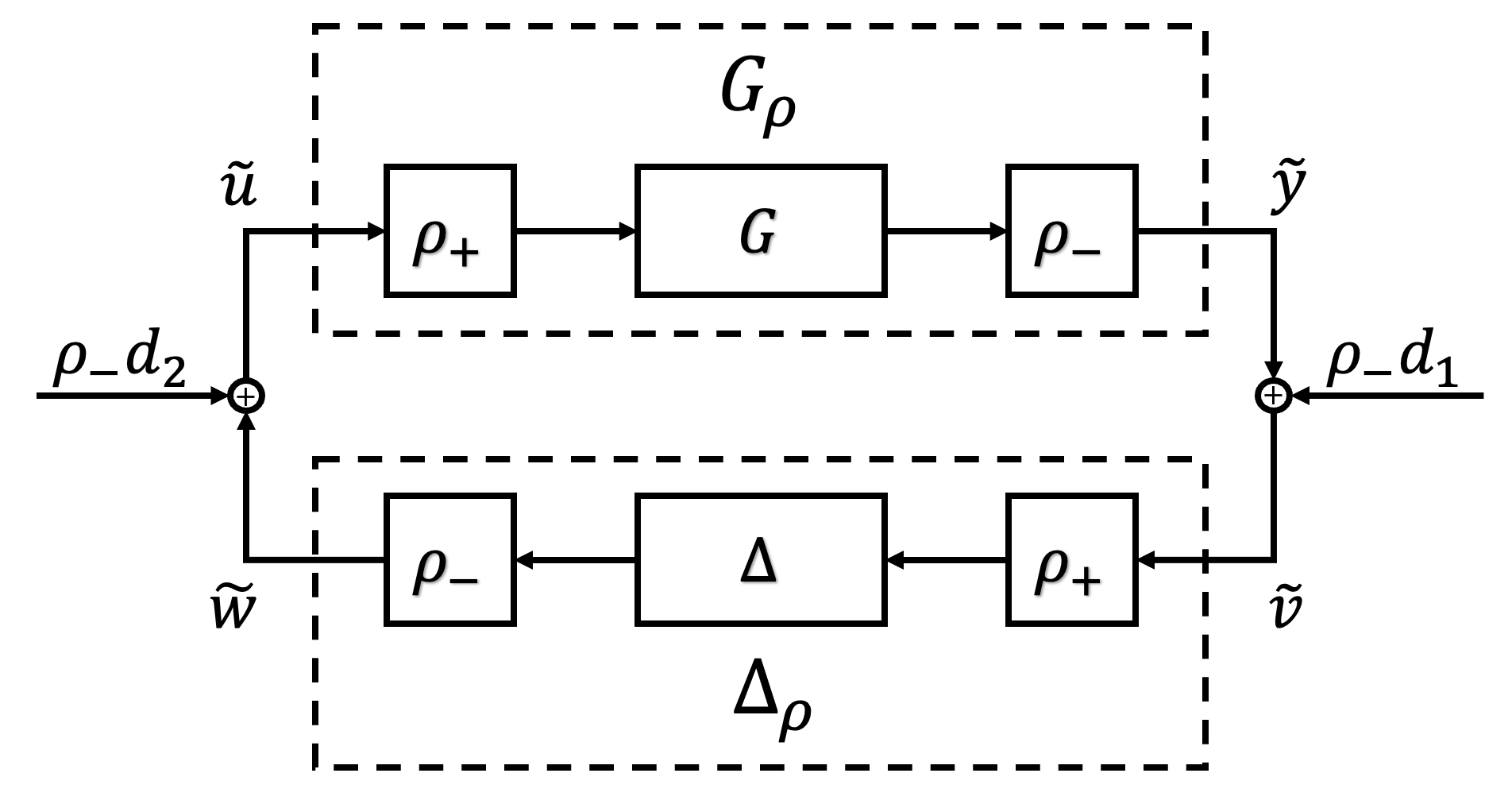}
    \caption{$\rho$-scaled feedback system}
    \label{fig:scaled lure}
\end{figure} 
{Consider the exponentially weighted feedback system in Fig.~\ref{fig:scaled lure} represented by
\begin{align}\label{eq:syst_eq_rho}
    \tilde{y}  =G_{\rho} \tilde{u};\quad  
    \tilde{w}  =\Delta_{\rho} \tilde{v};\quad
    \tilde{v}  =\tilde{y}+\rho_-d_1;\quad
    \tilde{u}  =\tilde{w} +\rho_-d_2
\end{align}} 
with $G_{\rho}=\rho_-G\rho_+$, $\Delta_{\rho}=\rho_-\Delta\rho_+$.
The two feedback systems $[G_\rho,\Delta_\rho]$ and $[G,\Delta]$ are related {in terms of well-posedness as well as feedback stability by the next lemma, which is established} via a similar manner to \cite[Theorem~2]{hu2016exponential}.

\begin{lemma}\label{lem:feedback_to_exponential_stable}
Suppose $G \in \RP$,  $\Delta:\ell_{2e}^{0+}\rightarrow \ell_{2e}^{0+}$, and  $[G,\Delta]$ in Fig.~\ref{fig:lure} is well-posed, then 
for $\rho\in(0,1]$, $[G_\rho,\Delta_\rho]$ in Fig.~\ref{fig:scaled lure} is well-posed. Additionally, $[G,\Delta]$  is exponentially stable with rate $\rho$ if $[G_\rho,\Delta_\rho]$ is stable.
\end{lemma}

\begin{proof}
 With the aid of Fig.~\ref{fig:scaled lure}, it is may be established that $(u,y,v,w)\in\ell_{2e}^{0+}$ is a solution to \eqref{eq:syst_eq} if and only if $(\tilde{u},\tilde{y},\tilde{v},\tilde{w})=(\rho_-u,\rho_-y,\rho_-v,\rho_-w) \in \ell_{2e}^{0+}$ is a solution to \eqref{eq:syst_eq_rho}. 
Hence the well-posedness of $[G_\rho,\Delta_\rho]$ is equivalent to that of $[G,\Delta]$.

Suppose that $[G,\Delta]$ and $[G_\rho,\Delta_\rho]$ are well-posed. For $\rho_-d_1,\rho_-d_2\in\ell_{2}^{0+}$, let $(u,y,v,w)$ and $(\tilde{u},\tilde{y},\tilde{v},\tilde{w})$ denote the  resultant solutions for   $[G,\Delta]$ with exogenous input $(d_1,d_2)\in \rho_{+}\ell_{2}^{0+}$ and $[G_\rho,\Delta_\rho]$ with exogenous input $(\rho_-d_1,\rho_-d_2)$ respectively. It follows that $\tilde{y}=\rho_- y$ and $\tilde{w} = \rho_- w$. The stability of $[G_\rho,\Delta_\rho]$ implies that there exists $\eta>0$ such that $\|\tilde{y}_k\|\le \eta$ and $\|\tilde{w}_k\|\le \eta$ for all $k\in\Z_0^{0+}$. Since 
\begin{align*}
\|\tilde{y}_k\|\le \eta  \text{  and  } \|\tilde{w}_k\|\le \eta 
 \Longleftrightarrow
\|y_k\|\le \eta \rho^{k} \text{  and  } \|w_k\|\le \eta\rho^{k}
\end{align*}
for all $ k\in\Z_0^{+}$, we have 
that $[G,\Delta]$ is exponentially stable with rate $\rho$ by Definition~\ref{def:robust exponential stability}. 
% the solutions to \eqref{eq:syst_eq} of $[G,\Delta]$ with external inputs $(d_1,d_2)$ are the same as those of $[G_\rho,\Delta_\rho]$ with $(d_1^\rho,d_2^\rho)$ as external inputs, where $d_1^\rho=\rho_-d_1$ and $d_2^\rho=\rho_-d_2$. Hence the well-posedness of $[G_\rho,\Delta_\rho]$ is equivalent to that of $[G,\Delta]$.
% With $d_1^\rho=\rho_-d_1$ and $d_2^\rho=\rho_-d_2$, it holds that $\tilde{y}=\rho_- y$ and $\tilde{w} = \rho_- w$ in Fig.~\ref{fig:scaled lure}. Thus,
% \begin{align*}
% \|y_k\|\le \eta \rho^{k} \text{  and  } \|w_k\|\le \eta\rho^{k}, \;\forall k\in\Z_0^{+} \Longleftrightarrow\\
% \|\tilde{y}_k\|\le \eta  \text{  and  } \|\tilde{w}_k\|\le \eta, \;\forall k\in\Z_0^{+}. 
% \end{align*}
% When $d_1,d_2\in\ell_2^{0+}$, the stability of $[G_\rho,\Delta_\rho]$ implies that $\lim_{k\to\infty}\|\tilde{y}_k\| = 0$ and $\lim_{k\to\infty}\|\tilde{w}_k\| = 0$,  whereby
% there exists $\eta>0$ such that
% $$\|y_k\|\le \eta \rho^{k} \text{  and  } \|w_k\|\le \eta\rho^{k}, \;\forall k\in\Z_0^{+}. $$
% Hence $[G,\Delta]$ is exponentially stable with rate $\rho$ by Definition~\ref{def:robust exponential stability}. 
\end{proof}
\color{black}

{We are now ready to present the proof of Theorem~\ref{th:fix rate}.}
\begin{proof}[{Proof of Theorem~\ref{th:fix rate}}]
In light of Lemma~\ref{lem:feedback_to_exponential_stable}, this theorem can be proved by showing  $[G_{\rho},\Delta_{\rho}]$ is robustly stable over $\Delta_\rho\in\mathbf{\Delta}_{\rho}$. In what follows, we show this by applying Proposition \ref{prop:SZ_IQC}. Let $\Delta_{\rho}$ be any element in $\mathbf{\Delta}_{\rho}$. We note the following observations.
\begin{enumerate}
    \item [(a)] From \ref{con:rho_homo_Delta}, there exists a mapping $\lambda\in[\underline{\lambda},\bar{\lambda}]\mapsto \Gamma_\lambda\in\mathbf{\Delta}_{\rho}$ that is continuous in the directed gap and satisfies $\Gamma_{\underline{\lambda}}=H$ and $\Gamma_{\bar{\lambda}} = \Delta_{\rho}$.
    \item [(b)]  From \ref{con:rho_Hstable}, $[G_{\rho},\Gamma_{\underline{\lambda}}]$ is stable.
    \item [(c)]  Since $\Gamma_\lambda\in\mathbf{\Delta}_{\rho}$ for $\lambda\in[\underline{\lambda},\bar{\lambda}]$, it can be obtained from \ref{con:rho_wellpose} and Lemma \ref{lem:feedback_to_exponential_stable} that $[G_\rho,\Gamma_{\lambda}]$ is well-posed for all $\lambda\in[\underline{\lambda},\bar{\lambda}]$.
    \item [(d)]  It follows from \ref{con:rho_IQC couple} and Lemma \ref{lem:graph_cf} that there exists $\Pi=\Pi^*\in\mathcal{L}_\infty$ such that  
    \begin{align*}
        \Bigg\langle \TwoOne{v}{w}, \Pi \TwoOne{v}{w}\Bigg\rangle\ge 0, \forall \TwoOne{v}{w}\in\mathscr{G}(\Gamma_{\lambda}), \forall \lambda\in[\underline{\lambda},\bar{\lambda}]
      \end{align*}
and 
         \begin{align*}
        \Bigg\langle \TwoOne{\tilde{y}}{\tilde{u}}, \Pi
        \TwoOne{\tilde{y}}{\tilde{u}}\Bigg\rangle \le -\epsilon \left\| \TwoOne{\tilde{y}}{\tilde{u}}\right\|^2, \forall \TwoOne{\tilde{y}}{\tilde{u}}\in\mathscr{G}'(G_\rho).
      \end{align*}
\end{enumerate}
{Matching (a)--(d) with conditions (i)--(iv) in Proposition~\ref{prop:SZ_IQC}, we obtain that $[G_\rho,\Gamma_\lambda]$ is stable for all $\lambda\in[\underline{\lambda},\bar{\lambda}]$. Since $\Gamma_{\bar{\lambda}} = \Delta_{\rho}$, we obtain that $[G_\rho,\Delta_\rho]$ is stable. Furthermore, as $\Delta_{\rho} \in \mathbf{\Delta}_{\rho}$ was arbitrary, it follows that $[G_\rho,\Delta_\rho]$ is robustly stable over $\Delta_\rho\in\mathbf{\Delta}_{\rho}$.}
\end{proof}

It is noteworthy that Theorem~\ref{th:fix rate} on external (input-output) stability with zero initial conditions is applicable to state-space systems for which internal stability with respect to arbitrary initial conditions may be established. This is particularly useful within the context of determining the speed of convergence of convex optimisation algorithms. To this end, let $G$ have a minimal realisation:
\begin{equation}\label{eq:ss_G}
G: \left\{\begin{array}{rl}
   \xi_{k+1} & =A_G \xi_{k}+B_G u_{k},\; \xi_0\in\R^{q} \\
    y_{k} & =C_G \xi_{k}
  \end{array}\right. .
\end{equation}

\begin{corollary}\label{cor:ss}
 Consider $G$ in \eqref{eq:ss_G} and $\Delta:\ell_{2e}^{0+}\rightarrow \ell_{2e}^{0+}$ belonging to $\mathbf{\Delta}$. Given $\rho\in(0,1)$,  the closed-loop system composed of \eqref{eq:ss_G} and $u=\Delta y$ satisfies  
 \begin{align}\label{eq:linear rate}
\lim_{k\rightarrow \infty} \rho^{-k}y_k = 0 
 \end{align}
for all $\xi_0\in\R^q$ and $\Delta\in\mathbf{\Delta}$  if the conditions in Theorem~\ref{th:fix rate} holds. 
\end{corollary}

\begin{proof} 
By Theorem~\ref{th:fix rate}, we have that $[G,\Delta]$ is robustly exponentially stable for all $\Delta\in\mathbf{\Delta}$ with decay rate $\rho$. 
Given any $\xi_0^*\in\R^q$, let $\tau>0$ and $\{u_k\}_{k=0,1,\cdots,\tau-1}$ be such that when fed to \eqref{eq:ss_G} with zero initial condition $\xi_0=0$ it results in $\xi_{\tau}=\xi_0^*$. Note that such a sequence $\{u_k\}_{k=0,1,\cdots,\tau-1}$ exists since $(A_G,B_G)$ is controllable. Let ${d_2}_k:=u_k$, ${d_1}_{k}:=-y_k$ for $k=0,1,\ldots,\tau-1$, and ${d_2}_k:=0$, ${d_1}_{k}:=0$ for all $\tau \le k\in\Z_0^+$. Obviously, $d_1,d_2\in \rho_+\ell_2^{0+}$. It follows from the robust exponential stability of $[G,\Delta]$ that $\rho_{-}y,\rho_{-}w\in\ell_2^{0+}$, and hence $\lim_{k\rightarrow \infty} \|\rho^{-k}y_k\|=0$. The claim follows since both $\Delta \in \mathbf{\Delta}$ and $\xi_0^* \in \R^q$ were arbitrary. 
\end{proof}

\subsection{An LMI characterisation} \label{sec: LMI}

When $\Pi$ is an element of $\RL$, the FDI in \eqref{eq:FDI_Grho} is a convex problem that can be verified via semidefinite programming, as is demonstrated next. Let $G$ have a minimal realisation $(A_G,B_G,C_G,0)$, we show in the following that the FDI in \eqref{eq:FDI_Grho} is equivalent to an LMI. To this end, first note from  \cite[Lemma~1]{Hu2017JRNC} that every self-adjoint $\Pi\in\RL$ can be factorised into 
\begin{align}\label{eq:Pi_factorisation}
 \Pi=\TwoOne{(e^{j\omega}I_{n}-A_{\Pi})^{-1}B_{\Pi}}{I}^*M_{\Pi} \TwoOne{(e^{j\omega}I_{n}-A_{\Pi})^{-1}B_{\Pi}}{I}
\end{align}
 with $A_\Pi$ being Schur stable.
\begin{proposition}\label{prop:equi_FDI_LMI}
Let $G\sim(A_G,B_G,C_G,0)$, ${\rho}\in(0,1)$, and $\Pi=\Pi^*\in\RL$ with a factorisation given in \eqref{eq:Pi_factorisation}. Then \eqref{eq:FDI_Grho} holds if and only if
there exists $P=P^{\top}$ such that 
\begin{align}\label{eq:LMI_prop9}
    \TwoOne{C^{\top}}{D^{\top}}M_{\Pi}\OneTwo{C}{D}+\TwoTwo{A^{\top}PA-P}{A^{\top} PB}{B^{\top} PA}{B^{\top} PB}<0,
\end{align} 
where
\begin{align*}
 A:=\TwoTwo{A_{\Pi}}{B_{\Pi}\TwoOne{C_G}{F}}{0}{\rho^{-1}(A_G+B_G F)},  & \; B:=\TwoOne{B_{\Pi}\TwoOne{0}{I}}{\rho^{-1}B_G} \\ C:=\TwoTwo{I_{n}}{0}{0}{\TwoOne{C_G}{F}} , & \;  D:=\TwoOne{0}{I},
\end{align*}
with   $F$ being any constant matrix such that $\rho^{-1}(A_G+B_G F)$ is Schur stable.
\end{proposition}
\begin{proof}
 By \cite[Theorem 5.6]{zhou1998essentials}, an RCF $(N_{{\rho}},M_{{\rho}})$ of $G_{{\rho}}\sim({{\rho}}^{-1}A_G,{{\rho}}^{-1}B_G,C_G,0)$ can be chosen to be
\begin{align*}
        \TwoOne{N_{{\rho}}}{M_{{\rho}}} & = \TwoOne{C_G}{F}[zI-{{\rho}}^{-1}(A_G+B_G F)]^{-1}({{\rho}}^{-1}B_G)+\TwoOne{0}{I} \\
         & =: \bar{C}(zI-\bar{A})^{-1}\bar{B}+\bar{D}, 
\end{align*}
where $F$ is any constant matrix such that ${\rho}^{-1}(A_G+B_G F)$ is stable and $\bar{A}:={\rho}^{-1}(A_G+B_G F)$, $\bar{B}:={\rho}^{-1}B_G$, 
$\bar{C}:=\mTwoOne{C_G}{F}$, $\bar{D}:=\mTwoOne{0}{I}$. The left-hand side of the FDI in \eqref{eq:FDI_Grho} is equal to 
\begin{align*}
  \left[\bar{C}(e^{j\omega}I-\bar{A})^{-1}\bar{B}+\bar{D}\right]^*  \TwoOne{(e^{j\omega}I_n-A_{\Pi})^{-1}B_{\Pi}}{I}^*M_{\Pi}\times  \quad \\\TwoOne{(e^{j\omega}I_n-A_{\Pi})^{-1}B_{\Pi}}{I}\left[\bar{C}(e^{j\omega}I-\bar{A})^{-1}\bar{B}+\bar{D}\right].
\end{align*}
Note that 
\begin{align*}
    & \TwoOne{(e^{j\omega}I_n-A_{\Pi})^{-1}B_{\Pi}}{I}\left[\bar{C}(e^{j\omega}I-\bar{A})^{-1}\bar{B}+\bar{D}\right] \\
   % = & \begin{bmatrix} I_n & 0 \\ 0 &\bar{C}
    %\end{bmatrix} \left(e^{j\omega}I-\TwoTwo{A_{\Pi}}{B_{\Pi}\bar{C}}{0}{\bar{A}}\right)^{-1}\TwoOne{B_{\Pi}\bar{D}}{\bar{B}}+\TwoOne%{0}{\bar{D}} \\
    = & \OneTwo{C}{D} \TwoOne {(e^{j\omega}I-A)^{-1}B}{I}.
\end{align*}
Therefore, the left-hand side of the FDI in \eqref{eq:FDI_Grho} can be expressed as
\begin{align*}
    \TwoOne {(e^{j\omega}I-A)^{-1}B}{I}^* \TwoOne{C^{\top}}{D^{\top}} M_{\Pi}  \OneTwo{C}{D}\TwoOne {(e^{j\omega}I-A)^{-1}B}{I}.
\end{align*}
Application of the discrete-time Kalman--Yakubovich--Popov (KYP) lemma~\cite{willems1971kyp} yields the equivalence between \eqref{eq:FDI_Grho} and the existence of $P=P^{\top}$ satisfying \eqref{eq:LMI_prop9}.
\end{proof}

% By the same argument in the proof above, one can also show that the FDI in \eqref{eq:FDI} is equivalent to the LMI in Proposition \ref{prop:equi_FDI_LMI} with $\rho=1$. 

\subsection{Illustrative examples} \label{sec: ex}

We demonstrate the usefulness of Theorem \ref{th:fix rate} with two examples. The following result for establishing the gap homotopy will be used in the examples.

\begin{proposition}\label{prop:gap_continuous_general}
Let $\Omega,\Gamma:\ell_{2e}^{0+}\rightarrow\ell_{2e}^{0+}$ be stable nonlinear systems  {and $K\in\mathcal{R}$}. Then the mapping $\beta\in[0,1] \mapsto F^\beta := K+(1-\beta)\Omega+\beta \Gamma$ is continuous in the directed gap $\vec{\delta}(\cdot,\cdot)$. 
\end{proposition}
\begin{proof}
The claim trivially holds if $\|\Omega-\Gamma\|= 0$. In what follows, suppose $\|\Omega-\Gamma\|\neq 0$. 
 Given $\beta\in[0,1]$, we have for any $d\neq 0$ satisfying $\beta+ d\in[0,1]$ that
\begin{align*}
    &\vec{\delta}(F^\beta,F^{\beta+d})=\limsup_{\tau\to\infty}\sup_{v\in\mathscr{G}_e (F^{\beta+d})}\inf_{\substack{u\in\mathscr{G}_e (F^\beta),\\ \|P_\tau u\|\neq 0}}\frac{\|P_\tau(u-v)\|}{\|P_\tau u\|} \\
    % &= \limsup_{\tau\to\infty}\sup_{y\in\ell_{2e}^{0+}}\inf_{\substack{x\in\ell_{2e}^{0+},\\ \|P_\tau x\|\neq 0}} \frac{\left\|P_\tau\left(\TwoOne{I}{F^{\beta} }x-\TwoOne{I }{F^{\beta+d}}y\right)\right\|}{\left\|P_\tau \TwoOne{I}{F^{\beta} }x\right\|} \\
    % &\leq \limsup_{\tau\to\infty}\sup_{\substack{y\in\ell_{2e}^{0+},\\ \|P_\tau y\|\neq 0}} \frac{\left\|P_\tau\left(\TwoOne{I}{F^{\beta} }y-\TwoOne{I}{F^{\beta+d}}y\right)\right\|}{\left\|P_\tau \TwoOne{I}{F^{\beta}}y\right\|} \\
    &\leq \limsup_{\tau\to\infty}\sup_{\substack{y\in\ell_{2e}^{0+},\\ \|P_\tau y\|\neq 0}} \frac{|d|\|P_\tau(\Omega-\Gamma)  y\|}{\left\|P_\tau y\right\|} \leq |d|\left\|\Omega-\Gamma\right\|.
\end{align*}%
For any $\epsilon>0$, let $\delta=\epsilon/\|\Omega-\Gamma\|>0$. Then, for all $\beta_1,\beta_2\in[0,1]$ satisfying $|\beta_1-\beta_2|<\delta$, it holds $\vec{\delta}(F^{\beta_1},F^{\beta_2})<\epsilon$. Thus, $\beta\in[0,1]\mapsto F^\beta$ is continuous in the directed gap.  
\end{proof}

{\begin{example}
Consider the LTI system $G(z)=1/(16z+2)$ and 
 nonlinear system $\Delta_{\alpha_1,\alpha_2}:u\in \ell_{2e}^{0+}\mapsto {y}\in\ell_{2e}^{0+}$ defined by \begin{align}\label{eq:example_Delta_ss}
x_{k+1} &= \frac{1}{2}\frac{x_k^3}{1+x_k^2} +u_k,~x_0 = 0,\\
y_k&=\alpha_1 x_k+\alpha_2 u_k \nonumber\end{align}
where $\alpha_1$ and $\alpha_2$ are uncertain parameters taking values in $[0,1]$. Define the set $\mathbf{\Delta}:=\{\Delta_{\alpha_1, \alpha_2}:~\alpha_1,\alpha_2\in[0,1]\}$. 
Our objectives are to investigate the robust exponential stability of the feedback system $[G,\Delta]$ over $\Delta\in\mathbf{\Delta}$ and to find an upper bound on the exponential decay rate by utilising Theorem \ref{th:fix rate}.     

Firstly, observe that $G\sim(-1/8,1,1/16,0)$ and $G_\rho \sim(-1/(8\rho),1/\rho,1/16,0)$. Let $H := I$, which is an element of $\mathbf{\Delta}$, we obtain that  the linear feedback system $[G_\rho,H]$ is stable for all $\rho\in(1/16,1]$. Thus, condition \ref{con:rho_Hstable} in Theorem \ref{th:fix rate} holds when $\rho\in(1/16,1]$. 

Secondly, since $H=I$ and $\mathbf{\Delta}_{\rho}=\{\alpha_1 \Delta_{1,0,\rho}+\alpha_2 H:~\alpha_1, \alpha_2\in[0,1]\}$, it follows that $H\in \mathbf{\Delta}_\rho$ for any $\rho\in(0,1]$. Next we show that given any $\Delta_\rho\in\mathbf{\Delta}_{\rho}$, there exists a mapping $\lambda\in[a,b]\mapsto \Gamma_\lambda, a<b$ continuous in gap such that $(\Gamma_a,\Gamma_b) =(H,\Delta_\rho)$. To this end, consider $a = \alpha_2$ and $b = 1+ \alpha_1$, and define the mapping by $\lambda\in[\alpha_2,1]\mapsto (1+\alpha_2-\lambda)H$ and $\lambda\in[1,1+\alpha_1]\mapsto (\lambda-1)\Delta_{0,1,\rho}+\alpha_2 H$. 
Applying Proposition~\ref{prop:gap_continuous_general} yields that $\lambda\in[\alpha_2,1+\alpha_1]\mapsto \Gamma_\lambda$ is continuous in directed  gap and thus $\mathbf{\Delta}_\rho$ is path-connected in the directed gap from $H$. Consequently, condition~\ref{con:rho_homo_Delta} of Theorem~\ref{th:fix rate} is satisfied.

Thirdly,  observe that $\gamma(G) = 0$, and therefore $[G,\Delta]$ is well-posed for all $\Delta\in\mathbf{\Delta}$, whereby \ref{con:rho_wellpose} of Theorem~\ref{th:fix rate} holds. 

Finally, we verify condition~\ref{con:rho_IQC couple} of Theorem~\ref{th:fix rate} with $\Pi=\mTwoTwo{5}{0}{0}{-1}$. Such a multiplier is associated with bounded systems; we show in what follows that $G_\rho$ and $\Delta_\rho$ satisfy the corresponding complementary IQCs defined by $\Pi$.  On the one hand, $\|G_\rho\|_{\infty} = 1/(16\rho-2)$ for $\rho\in(1/8,1]$, and thus $G_\rho$ satisfies the FDI \eqref{eq:FDI_Grho} for $\rho\in(7/16,1)$. On the other hand, 
the state-space representation of $\Delta_{\alpha_1,\alpha_2,\rho} = \rho_-\Delta_{\alpha_1,\alpha_2}\rho_+:~\tilde{u}\in\ell_{2e}^{0+}\mapsto \tilde{y}\in\ell_{2e}^{0+}$, given by 
\begin{align*}\label{eq:example_DeltaRho_ss}\tilde{x}_{k+1} & = \frac{\rho^{-1}}{2}\frac{\tilde{x}_k^3}{\rho^{-2k}+\tilde{x}_k^2} +\rho^{-1}\tilde{u}_k ,~\tilde{x}_0=0,\\
\tilde{y}_{k} & = \alpha_1 \tilde{x}_k+\alpha_2 \tilde{u}_k,
\end{align*}
satisfies $\|\tilde{y}\| \leq 2 \alpha_1\|\tilde{u}\|/\sqrt{2\rho^2-1}+\alpha_2\|\tilde {u}\|\le (2 /\sqrt{2\rho^2-1}+1)\|\tilde{u}\|$ for $\rho\in(\sqrt{2}/2,1]$, whereby $\Gamma_\rho$ is bounded with gain $\|\Gamma_\rho\| \leq 2/\sqrt{2\rho^2-1}+1$. 
% Note that $\|(1-\beta)\Gamma_\rho+\beta I\|\le (1-\beta)\|\Gamma_\rho\|+\beta\le \max\{1,2/\sqrt{2\rho^2-1}\}$ for all $\beta\in[0,1]$. 
Hence, $\Delta_\rho\in\mathbf{\Delta}_\rho$ satisfies IQC \eqref{eq:rho_IQC on Delta} defined by $\Pi$ whenever $\rho>\sqrt{5/8}$. Combining all the constraints on $\rho$ above, we obtain $\rho > \max\{1/16, 1/8,7/16,\sqrt{2}/2, \sqrt{5/8} \} \approx 0.7906$. To conclude, conditions \ref{con:rho_Hstable} to \ref{con:rho_IQC couple} in Theorem \ref{th:fix rate} hold whenever $\rho >0.7906$, so a bound on the robust exponential decay rate for $[G,\Delta]$ over $\Delta\in\mathbf{\Delta}$ may be obtained as $0.7906$ via Theorem~\ref{th:fix rate}.
\end{example}
}

The example above is concerned with an uncertain feedback interconnection of open-loop bounded systems. An uncertain feedback interconnection of time-varying open-loop unbounded systems is considered in the next example.

{\begin{example}
Consider the unstable LTI system  $G(z)= (3-6z)/ (z-2)$ and unstable time-varying system $\Delta$ given by
 \begin{align*}
x_{k+1}&=\lambda^{-1}2x_k+\lambda^{-1}u_k,\,x_0=0\\
     y_k&=-9x_k-d_ku_k,
 \end{align*}
 where $\lambda\in[1,2]$ is parametric uncertainty and $d_k\in[5,7]$ is a time-varying signal. Observe that such an uncertain system can be alternatively characterised by the uncertainty set $\mathbf{\Delta} :=\{H_\lambda + S:~\|S\| \leq 1,~\lambda\in[1,2]\}$ with $H(z) = (3-6z)/ (z-2)=G(z)$ and $S:~{u}\in\ell_{2e}^{0+}\mapsto {y}\in\ell_{2e}^{0+}$ being any static sector bounded nonlinearity in $[-1,1]$. Assume that the feedback system $[G,\Delta]$ is well-posed for all $\Delta \in \mathbf{\Delta}$.
The objectives here are to show that the feedback system $[G,\Delta]$ is robustly exponentially stable over $\Delta\in\mathbf{\Delta}$ and to find an upper bound on the exponential decay rate by utilising Theorem \ref{th:fix rate}.     

Firstly,  one can verify that  the linear feedback system $[G,H]$ is stable. Let
\[
\rho_1^\star:=\inf\{\rho_1\in(0,1]:~[G_\rho,H]~\text{is stable},~\forall~\rho\in(\rho_1,1]\}.
\]
Note that $\rho_1^\star < 1$ by continuity.
 Thus, condition \ref{con:rho_Hstable} in Theorem \ref{th:fix rate} holds whenever $\rho\in(\rho_1^\star,1]$. 

Secondly, since $H=G$ and $\mathbf{\Delta}_{\rho}=\{H_{\lambda\rho} + S_\rho:~\|S\| \leq 1,~\lambda\in[1,2]\}$,  one may verify that $H\in \mathbf{\Delta}_\rho$ for any $\rho\in(1/2,1]$. Now let $\rho\in(1/2,1]$, we show next that condition~\ref{con:rho_homo_Delta} of Theorem~\ref{th:fix rate} is satisfied i.e., for any $\Delta_\rho\in\mathbf{\Delta}_\rho$, there exists a mapping $\lambda\in[a,b]\mapsto\Gamma_\lambda$, $a<b$ that is continuous in the directed gap with $(\Gamma_a,\Gamma_b)=(H,\Delta_\rho)$. Consider $\Delta_\rho = H_{\lambda\rho}+S_\rho \in \mathbf{\Delta}_\rho$ and note that $\lambda\rho\in(1/2,2]$.
When $\lambda\rho < 1$, define $\Gamma_\eta := H_{1-\eta}$ for $\eta\in[0,1-\lambda\rho]$ and $\Gamma_\eta := H_{\lambda\rho}+(\eta - 1 + \lambda\rho) S_\rho$ for $\eta \in (1 - \lambda\rho, 2 - \lambda\rho]$, whereby $\Gamma_0 = H$ and $\Gamma_{2 -\lambda\rho} = \Delta_\rho$.
% For $\rho \in (1/2, 1]$ and $\lambda \in [1, 2]$, consider $\Delta_\rho = H_{\lambda\rho}+S_\rho \in \mathbf{\Delta}_\rho$. When $\lambda\rho \leq 1$, define $\Gamma_\eta := H_{1-\eta}$ for $\eta\in[0,1-\lambda\rho]$ and $\Gamma_\eta := H_{\lambda\rho}+(\eta - 1 + \lambda\rho) S_\rho$ for $\eta \in (1 - \lambda\rho, 2 - \lambda\rho]$, whereby $\Gamma_0 = H$ and $\Gamma_{2 -\lambda\rho} = \Delta_\rho$. 
Note that $\eta \in [0,1-\lambda\rho] \mapsto \Gamma_\eta$ is continuous in the directed gap and $\eta \in [1 - \lambda\rho, 2 - \lambda\rho] \mapsto \Gamma_\eta$ is continuous in the directed gap by Proposition~\ref{prop:gap_continuous_general} with $(K,\Omega,\Gamma) = (H_{\lambda\rho},0,S_\rho)$. When $\lambda\rho \ge 1$, a continuous path can be constructed similarly. 

Finally, we verify condition~\ref{con:rho_IQC couple} of Theorem~\ref{th:fix rate} with $\Pi = \mTwoTwo {-1} {0}{0}  {1}$. One may observe that the IQC defined by a multiplier of this form enforces a minimum gain on the system, which is well suited for certain unbounded systems. We show in what follows that $G_\rho$ and $\Delta_\rho$ satisfy the corresponding complementary IQCs defined by $\Pi$. To this end,
note that $|G(e^{j\omega})|=3$, $\forall\omega\in[0,2\pi]$. Let 
\[
\rho_2^\star:=\inf\{\rho_2\in(0,1]:~|G_\rho(e^{j\omega})|>1,~\forall~\rho\in(\rho_2,1],~\omega\in[0,2\pi]\},
\]
in which case for all $\rho\in(\rho_2^\star,1]$, $G_\rho$ satisfies the FDI \eqref{eq:FDI_Grho} with $\Pi$. Similarly, note that $|H(e^{j\omega})|=3$, $\forall\omega\in[0,2\pi]$ and let
\[
\rho_3^\star:=\inf\{\rho_3\in(0,1]:~\tau(H_{\lambda\rho})>2,~\forall~\rho\in(\rho_3,1],~\lambda\in[1,2]\},
\]
where $\tau(\cdot)$ is defined for any $X:\ell_{2e}^{0+}\to \ell_{2e}^{0+}$ by
\[
\tau(X):=\inf_{\substack{u\in\ell_{2e}^{0+},\\ P_ku \neq 0,k \in \mathbb{Z}_0^+}}\dfrac{\|P_kXu\|}{\|P_ku\|}. 
\] 
Observe that $\|S_\rho\| \leq 1$ as $S$ is a static nonlinearity satisfying $\|S\| \leq 1$. Consequently, for $\rho\in(\rho_3^\star,1]$, it holds $\tau(\Delta_\rho) \geq \tau(H_{\lambda\rho})-\|S_\rho\| > 2-1 =1$.
Hence, $\Delta_\rho\in\mathbf{\Delta}_\rho$ satisfies IQC \eqref{eq:rho_IQC on Delta} with $\Pi$ whenever $\rho\in(\rho_3^\star,1]$. 

Combining all the constraints on $\rho$ above, we obtain that whenever $\rho > \rho^\star:=\max\{\rho_1^\star,\rho_2^\star,\rho_3^\star, 1/2\}$, conditions \ref{con:rho_Hstable} to \ref{con:rho_IQC couple} in Theorem~\ref{th:fix rate} hold. A bisection search for the optimal $\rho_i^\star$, $i=1,2,3$ yields that 
$$\rho_1^\star=0.412,~\rho_2^\star=0.714,~\rho_3^\star=0.875,$$
whereby $\rho^\star = 0.875$. To conclude, a bound on the robust exponential decay rate for $[G,\Delta]$ over $\Delta\in\mathbf{\Delta}$ can be obtained as $0.875$ by Theorem~\ref{th:fix rate}.
\end{example}
}

\color{black}

\subsection{Lurye systems with static nonlinearity} \label{subsec: Lurye}

In this subsection, we specialise Theorem~\ref{th:fix rate}  to  Lurye systems $[G, \Delta]$, where $G \in \mathcal{R}$ is possibly unstable and has a minimal realisation $(A_G,B_G,C_G,0)$, and $\Delta$ belongs to a nonlinearity class that is static, bounded and locally Lipschitz continuous. To be more specific, a system $\Delta:\ell_{2e}^{0+}\rightarrow\ell_{2e}^{0+}$ is said to be static (a.k.a. memoryless) if there exists $\phi:\R^p\rightarrow\R^p$ such that $(\Delta(v))_k=\phi(v_k)$ for all $v\in\ell_{2e}^{0+}$ and $k\in\Z_0^+$.

A version of Theorem~\ref{th:fix rate} for Lurye systems is warranted. 
\begin{theorem}\label{thm:main for Lurye}
Consider the feedback system in Fig.~\ref{fig:lure} with $G,\Delta:\ell_{2e}^{0+}\rightarrow \ell_{2e}^{0+}$. Suppose that $G \sim (A_G, B_G, C_G, 0)$ and $\Delta\in\mathbf{\Delta}$ is static and bounded. 
{Given $\rho\in(0,1)$, the feedback system $[G,\Delta]$ is  robustly exponentially stable over $\Delta\in\mathbf{\Delta}$ with decay rate $\rho$,} if the following conditions are satisfied:
\begin{enumerate} \renewcommand{\theenumi}{\textup{(\roman{enumi})}}\renewcommand{\labelenumi}{\theenumi} 
\item \label{cond:homoDelta} there exists {$c\in\mathbb{R}$} such that $(1 - \beta) \Delta + \beta c I \in \mathbf{\Delta}$ for all $\Delta \in \mathbf{\Delta}$, $\beta \in [0, 1]$;

\item\label{cond:stab_initial} $r(A_G+cB_GC_G)< \rho$;

\item \label{cond:Lurye(3)} there exists  $\Pi=\Pi^*\in\mathcal{L}_\infty$ such that 
{\begin{align}\label{eq:Pi_Delta}
 \Bigg\langle \TwoOne{\tilde{v}}{\tilde{w}}, \Pi \TwoOne{\tilde{v}}{\tilde{w}}\Bigg\rangle\ge 0, \forall \TwoOne{\tilde{v}}{\tilde{w}}\in\mathscr{G}(\Delta_\rho), \Delta\in\mathbf{\Delta}, 
 \end{align}}
and 
\begin{align} \label{eq:th2_FDI}  
\TwoOne{N_{\rho}(e^{j\omega})}{M_{\rho}(e^{j\omega})}^{*}\Pi(e^{j\omega})\TwoOne{N_{\rho}(e^{j\omega})}{M_{\rho}(e^{j\omega})}<0, \forall \omega\in[0,2\pi],
\end{align}
where $(N_{\rho},M_{\rho})$ is an RCF of $G_{\rho}$. 
\end{enumerate}
% In addition, given $\rho\in(0,1)$, the feedback system $[G,\Delta]$ is  robustly exponentially stable over $\Delta\in\mathbf{\Delta}$ with decay rate $\rho$, if \ref{cond:homoDelta} above holds, $r(A_G+cB_GC_G)< \rho$, and there exists  $\Pi=\Pi^*\in\mathcal{L}_\infty$ such that \eqref{eq:Pi_Delta} holds 
% and 
% \begin{align} \label{eq:th2_FDI}  
% \TwoOne{N_{\rho}(e^{j\omega})}{M_{\rho}(e^{j\omega})}^{*}\Pi(e^{j\omega})\TwoOne{N_{\rho}(e^{j\omega})}{M_{\rho}(e^{j\omega})}<0, \forall \omega\in[0,2\pi],
% \end{align}
% where $(N_{\rho},M_{\rho})$ is an RCF of $G_{\rho}$. 
\end{theorem}

\begin{proof}
Firstly, note that if $r(A_G+cB_GC_G)< \rho$, then $[G_\rho,cI]$ is stable, whereby condition~\ref{con:rho_Hstable} in Theorem~\ref{th:fix rate} is satisfied. 

Secondly, for any $\rho\in(0,1]$, recall that $\mathbf{\Delta}_\rho := \{\Delta_\rho:~\Delta\in\mathbf{\Delta}\}$. Since $\Delta\in\mathbf{\Delta}$ is static and bounded, it follows from~\cite[Appendix~A.2]{michalowsky2021robust} that $\Delta_\rho$ is bounded for any $\rho\in(0,1]$. By \ref{cond:homoDelta}, we have for every $\Delta_\rho\in\mathbf{\Delta}_\rho$ and $\beta\in[0,1]$ that $(1-\beta)\Delta_\rho+\beta cI\in\mathbf{\Delta}_\rho$. Define the mapping 
$\beta\in[0,1]\mapsto \Delta_\rho^\beta = (1-\beta)\Delta_\rho+\beta cI\in\mathbf{\Delta}_\rho,$
which is continuous in the directed gap by Proposition~\ref{prop:gap_continuous_general} with $(K,\Omega,\Gamma) = (0,\Delta_\rho,cI)$. In other words, $\mathbf{\Delta}_\rho$ is path-connected from $cI$ in the directed gap for each $\rho\in(0,1]$. Therefore, condition~\ref{con:rho_homo_Delta} in Theorem \ref{th:fix rate} is satisfied.  

 Thirdly, note that $G \sim (A_G, B_G, C_G, 0)$ is strictly causal, whereby its uniform instantaneous gain is zero, i.e., $\gamma(G)=0$.  It then follows from $\gamma(G)\gamma(\Delta)<1$ and Lemma~\ref{lem:wellposed_inst} that $[G,\Delta]$ is well-posed for all $\Delta\in\mathbf{\Delta}$. Consequently, condition \ref{con:rho_wellpose} in Theorem \ref{th:fix rate} is satisfied.  
 
Fourthly, by \eqref{eq:Pi_Delta} and \eqref{eq:th2_FDI}, condition~\ref{con:rho_IQC couple} in Theorem~\ref{th:fix rate} is satisfied.

Combining the analyses above, we obtain from Theorem~\ref{th:fix rate} with $H=cI$ that the feedback system $[G,\Delta]$ is robustly exponentially stable over $\Delta\in\mathbf{\Delta}$ with the given decay rate $\rho$. 
\end{proof}

It is worth noting that Theorem~\ref{thm:main for Lurye} is a specialisation of Theorem~\ref{th:fix rate} to Lurye systems, {which does not} follow trivially from Proposition~\ref{prop:SZ_IQC}.

As a case study, we consider below Lurye systems with sector-bounded nonlinearity. 
Given $a\le b$, the static nonlinearity $\Delta$ is said to be $(a,b)$-sector-bounded if  for all $x\in\R^p$, 
\begin{align*}
    (bx-\phi (x))^\top(\phi (x)-ax)\ge 0.
\end{align*}
We denote the class of $(a,b)$-sector-bounded  nonlinearities by $\mathbf{\Delta}_{a,b}$. 
For $\mathbf{\Delta}=\mathbf{\Delta}_{a,b}$, condition \ref{cond:homoDelta} in Theorem \ref{thm:main for Lurye} can be easily verified. Specifically,  for every $\Delta\in\mathbf{\Delta}_{a,b}$, $c\in[a,b]$ and $\beta\in[0,1]$, it holds that
$(1-\beta)\Delta+\beta cI\in\mathbf{\Delta}_{a,b}$. 
Letting  
\[
\Pi_{a,b}=\TwoTwo{-2ab}{a+b}{a+b}{-2},
\]
it is well-known (from e.g. \cite[Section VI]{megretski1997system}) that 
\begin{align}
            \Bigg\langle \TwoOne{v}{w}, \Pi_{a,b} \TwoOne{v}{w}\Bigg\rangle\ge 0, \forall \TwoOne{v}{w}\in\mathscr{G}(\Delta),\,\Delta\in\mathbf{\Delta}_{a,b}.  \label{eq:IQC on sector_Delta_rho}
\end{align}
Moreover,  since $\Delta\in \mathbf{\Delta}_{a,b}$ is static, it holds that
\[\bigcup_{\Delta\in\mathbf{\Delta}_{a,b}}\mathscr{G}(\Delta) = \bigcup_{\Delta\in\mathbf{\Delta}_{a,b}} \mathscr{G}(\Delta_\rho),\,\forall \rho>0.\]
Thus, \eqref{eq:Pi_Delta} in condition \ref{cond:Lurye(3)} of Theorem \ref{thm:main for Lurye} holds with $\Pi=\Pi_{a,b}$ for any $\rho>0$.
To apply Theorem \ref{thm:main for Lurye}, what remain to be verified are condition \ref{cond:stab_initial} and  \eqref{eq:th2_FDI}. By employing the LMI characterisation in Proposition \ref{prop:equi_FDI_LMI}, we obtain the following efficiently computable conditions. 
\begin{proposition}\label{th:LMI_sector}
    Consider a Lurye system  in Fig. \ref{fig:lure} with $G,\Delta:\ell_{2e}^{0+}\rightarrow \ell_{2e}^{0+}$. Suppose that $G\sim(A_G,B_G,C_G,0)$, $\Delta$  belongs to $\mathbf{\Delta}_{a,b}$ for some $a \le b$.  Given $\rho\in(0,1)$, the feedback system $[G,\Delta]$ is robustly exponentially stable with rate $\rho$ for all $\Delta\in\mathbf{\Delta}_{a,b}$ if there exists $c\in[a,b]$ such that
\begin{align}\label{eq:con Ga expstable}
r(A_G+ cB_GC_G) \le \rho
\end{align} 
and 
there exists $P=P^\top$ such that
 \begin{align}\label{eq:LMI_sector}
\TwoTwo{\bar{C}^\top\Pi_{a,b}\bar{C}}{\bar{C}^\top\Pi_{a,b}\bar{D}}{\bar{D}^\top\Pi_{a,b}\bar{C}}{\bar{D}^\top\Pi_{a,b}\bar{D}} + \TwoTwo{\bar{A}^\top P \bar{A}-P}{\bar{A}^\top P\bar{B}}{\bar{B}^\top P\bar{A}}{\bar{B}^\top P\bar{B}}<0
\end{align}
where
\begin{align}
 \bar{A}:=\rho^{-1}(A_G+B_G F),  & \; \bar{B}:=\rho^{-1}B_G, \nonumber \\ \bar{C}:=\TwoOne{C_G}{F}, & \;  \bar{D}:=\TwoOne{0}{I} \label{eq:ABCDbar}
\end{align}
and $F$ is any constant matrix such that $\rho^{-1}(A_G+B_G F)$ is stable.
\end{proposition}
\begin{figure}
    \centering
    \includegraphics[scale=0.28]{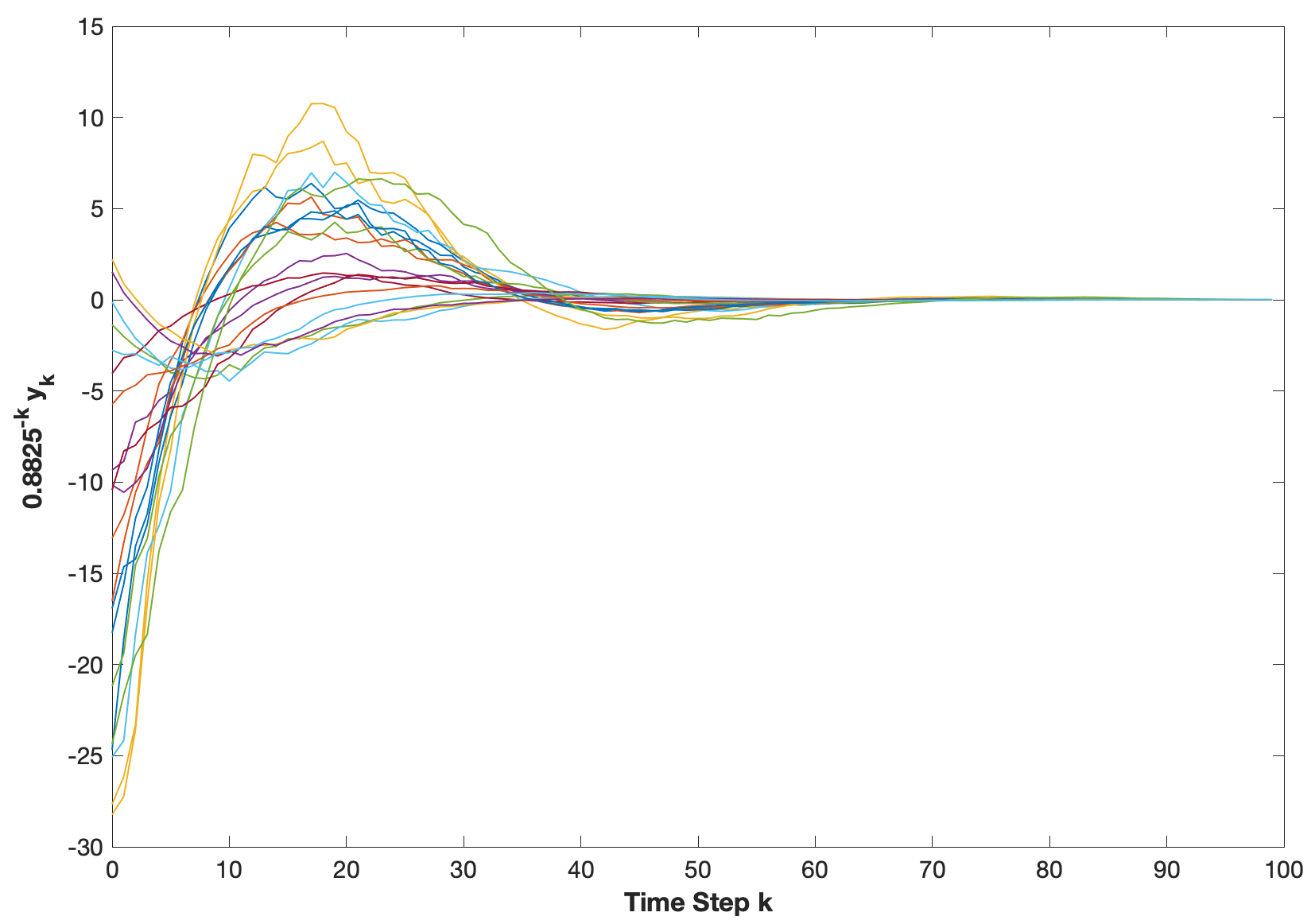}
    \caption{The trajectories of $0.8825^{-k}y_k$ under various initial conditions}
    \label{fig:ex_sector}
\end{figure}

\begin{example}Consider  the closed-loop system \eqref{eq:ss_G} with  $G\sim (A_G, B_G, C_G, 0) = (\mTwoTwo{1.1}{0}{0}{0.9},\mTwoOne{1}{1},\mOneTwo{-3}{0.5},0)$, and $u_k=\mathrm{X}_ky_k$ where $\mathrm{X}_k, k\in\Z_0^+$ is a uniformly distributed sequence in the interval $[a, b] = [0.1,0.2]$. 
Note that $G$ is unstable since $r(A_G)=1.1$ and the set $\{(u,y)\in\ell_2^{0+} :~u_k=\mathrm{X}_ky_k\}$ can be characterised by  $\bigcup_{\Delta\in\mathbf{\Delta}_{0.1,0.2}}\mathscr{G}(\Delta)$. By exploiting Proposition \ref{th:LMI_sector}, we obtain below the upper bound of the exponential rate at which $y$ decays to $0$ under all initial conditions of $G$. To this end, first observe that $r(A_G+aB_GC_G)=0.8803$, so from condition \eqref{eq:con Ga expstable} we know that $\rho\in[0.8803,1]$. Then, by solving the LMI problem in \eqref{eq:LMI_sector} while minimising $\rho$ (e.g., by running a bisection algorithm over $\rho$ and at each step performing the LMI feasibility test for fixed $\rho$), we obtain that the optimal $\rho$ is given by $\rho^*=0.8825$. Hence, the closed-loop system with $G$ and $u_k=\mathrm{X}_ky_k$ satisfies that $\lim_{k\rightarrow\infty}{\rho^{*}}^{-k}y_k=0$ for any initial conditions $\xi_0\in\R^2$. Under various randomly generated initial conditions and $X_k$, the trajectories of $\rho^*_-y$ are illustrated in Fig.~\ref{fig:ex_sector}. 
\end{example}
\color{black}

\section{Conclusion} \label{sec: con}
We derived conditions under which a feedback interconnection of possibly unbounded and uncertain open-loop systems achieves both robust closed-loop stability and exponential convergence in the endogenous signals. The conditions are expressed in terms of a homotopy in the directed gap for exponentially weighted systems and integral quadratic constraints along the homotopy. When specialised to Lurye systems, where the nonlinearity is static and bounded, the conditions, {in particular the homotopy requirement}, are shown to be considerably easier to satisfy and verify. Future research directions of interest include the consideration of linear time-varying dynamics and alternative weightings for establishing sub-exponential and super-exponential convergence rates.

 \bibliography{ZF_Opt_TAC.bib}
\end{document}